\documentclass[onecolumn,showpacs,fleqn,nobibnotes]{revtex4}

\usepackage{amsmath}
\usepackage{graphicx}
\usepackage{float}
\usepackage{subfigure}

\def\lsim{\raise0.3ex\hbox{$<$\kern-0.75em\raise-1.1ex\hbox{$\sim$}}}
\def\gsim{\raise0.3ex\hbox{$>$\kern-0.75em\raise-1.1ex\hbox{$\sim$}}}

\newcommand{\beq}{\begin{equation}}
\newcommand{\eeq}{\end{equation}}

\begin{document}

\title{Quarkonium+$\gamma$ production in coherent hadron - hadron  interactions at LHC energies}
\author{V.P. Gon\c{c}alves$^{a}$ and M.M. Machado$^{b}$}

\affiliation{$^{a}$ Instituto de F\'{\i}sica e Matem\'atica, Universidade Federal de
Pelotas\\
Caixa Postal 354, CEP 96010-900, Pelotas, RS, Brazil.}

\affiliation{$^{b}$ Instituto Federal de Educa\c{c}\~ao, Ci\^encia e Tecnologia, IF - Farroupilha,
Campus S\~ao Borja \\ Rua Otaviano Castilho Mendes, 355, CEP 97670-000 - S\~ao Borja, RS, Brazil.}

\begin{abstract}
In this paper we study the $H + \gamma$ ($H = J/\Psi$ and $\Upsilon$) production in coherent hadron - hadron interactions at  LHC energies. Considering the ultrarelativistic protons as a source of photons, we estimate the $\gamma + p \rightarrow H + \gamma + X$ cross section  using the non-relativistic QCD (NRQCD) factorization formalism and considering different sets of values for the matrix elements. Our results  for the total $ p + p \rightarrow p + H + \gamma + X$ cross sections and rapidity distributions at $\sqrt{s} = 7$ and 14 TeV demonstrate that the experimental analysis of the $J/\Psi + \gamma$ production at LHC is feasible.

\end{abstract}
\pacs{12.40.Nn, 13.85.Ni, 13.85.Qk, 13.87.Ce}
\maketitle

\section{Introduction}

In the last years, the analysis of coherent hadron-hadron collisions becomes  an alternative way to study the theory of strong interactions - the Quantum Chromodynamics (QCD) - in the regime of high energies (For  reviews see Ref. \cite{upc}).
The basic idea in coherent  hadron collisions is that
the total cross section for a given process can be factorized in
terms of the equivalent flux of photons { in} the hadron projectile and
the photon-photon or photon-hadron production cross section.
The main advantage of using colliding hadrons and nuclear beams for
studying photon induced interactions is the high equivalent photon
energies and luminosities, that can be achieved at existing and
future accelerators.
 Consequently, studies of $\gamma p$ interactions
at LHC  provides valuable information on the QCD dynamics at
high energies. The photon-hadron interactions can be divided into exclusive and inclusive reactions. In the first case, a certain particle is produced, while the target remains in the ground state (or is only internally excited). On the other hand, in inclusive interactions the particle produced is accompanied by one or more particles from the breakup of the target. The typical examples of these processes are the exclusive vector meson production, described by the process $\gamma p \rightarrow H p$ ($H =  J/\Psi, \Upsilon$), and the inclusive heavy quark production [$\gamma p \rightarrow X Y$ ($X = c\overline{c}, b\overline{b}$)], respectively. The results of these studies demonstrate that their detection is feasible at the LHC (For recent discussions see, e.g. Refs. \cite{vicmairon,vicmag_update,vicwerner}). It motivates the analysis of the production of other final states in coherent hadron - hadron interactions.

In this paper we study, for the first time, the inclusive quarkonium + photon photoproduction in $pp$ collisions at LHC energies. The total cross section for the process $ p + p \rightarrow p \otimes H + \gamma + X$ ($H = J/\Psi$ or $\Upsilon$) is given by
\begin{eqnarray}
\sigma (p p \rightarrow p \otimes H + \gamma + X) =  2 \int  d\omega  \,\frac{dN_{\gamma/p}(\omega)}{d\omega}\,\sigma_{\gamma p \rightarrow H + \gamma + X} (W_{\gamma p}^2) \,,
\label{sighh}
\end{eqnarray}
where $\otimes$ represents a rapidity gap in the final state, $\omega$ is the photon energy,   $\frac{dN_{\gamma}}{d\omega}$ is the equivalent photon flux, $W_{\gamma p}^2=2\,\omega \sqrt{S_{\mathrm{NN}}}$  and $\sqrt{S_{\mathrm{NN}}}$ is  the c.m.s energy of the
hadron-hadron system.
The  photon spectrum of a relativistic proton is given by  \cite{Dress},
\begin{eqnarray}
\frac{dN_{\gamma/p}(\omega)}{d\omega} =  \frac{\alpha_{\mathrm{em}}}{2 \pi\, \omega} \left[ 1 + \left(1 -
\frac{2\,\omega}{\sqrt{S_{NN}}}\right)^2 \right]
\left( \ln{\Omega} - \frac{11}{6} + \frac{3}{\Omega}  - \frac{3}{2 \,\Omega^2} + \frac{1}{3 \,\Omega^3} \right) \,,
\label{eq:photon_spectrum}
\end{eqnarray}
with the notation $\Omega = 1 + [\,(0.71 \,\mathrm{GeV}^2)/Q_{\mathrm{min}}^2\,]$, $Q_{\mathrm{min}}^2= \omega^2/[\,\gamma_L^2 \,(1-2\,\omega /\sqrt{S_{NN}})\,] \approx (\omega/
\gamma_L)^2$ and $\gamma_L$ is the Lorentz boost  of a single beam.
Distinctly from the exclusive vector meson photoproduction discussed in, e.g., Ref. \cite{vicmag_update}, which is characterized by two rapidity gaps in the final state,  in our case it is characterized by one rapidity gap associated to the photon emitted by one of the protons and the remnants of the other proton.
The main input in our calculations is the $\gamma + p \rightarrow H + \gamma + X$ cross section, which we estimate using the non-relativistic QCD (NRQCD) factorization formalism \cite{nrqcd}. Recent results demonstrate that this formalism is able to describe quite well the RHIC and HERA data for the  hadroproduction and photoproduction of charmonium in terms of an universal set of matrix elements and that the inclusion of the color octet processes is indispensable to describe the photoproduction data \cite{bute}. 

This paper is organized as follows. In next section we present a brief review about the quarkonium+$\gamma$ photoproduction in the NRQCD formalism. In Section \ref{results} we present our predictions for the rapidity distributions and total cross sections for $J/\Psi + \gamma$ and $\Upsilon + \gamma$ production at LHC energies. Finally, in Section \ref{conc}, we summarize our main conclusions.

\begin{figure}
\includegraphics[scale=0.2]{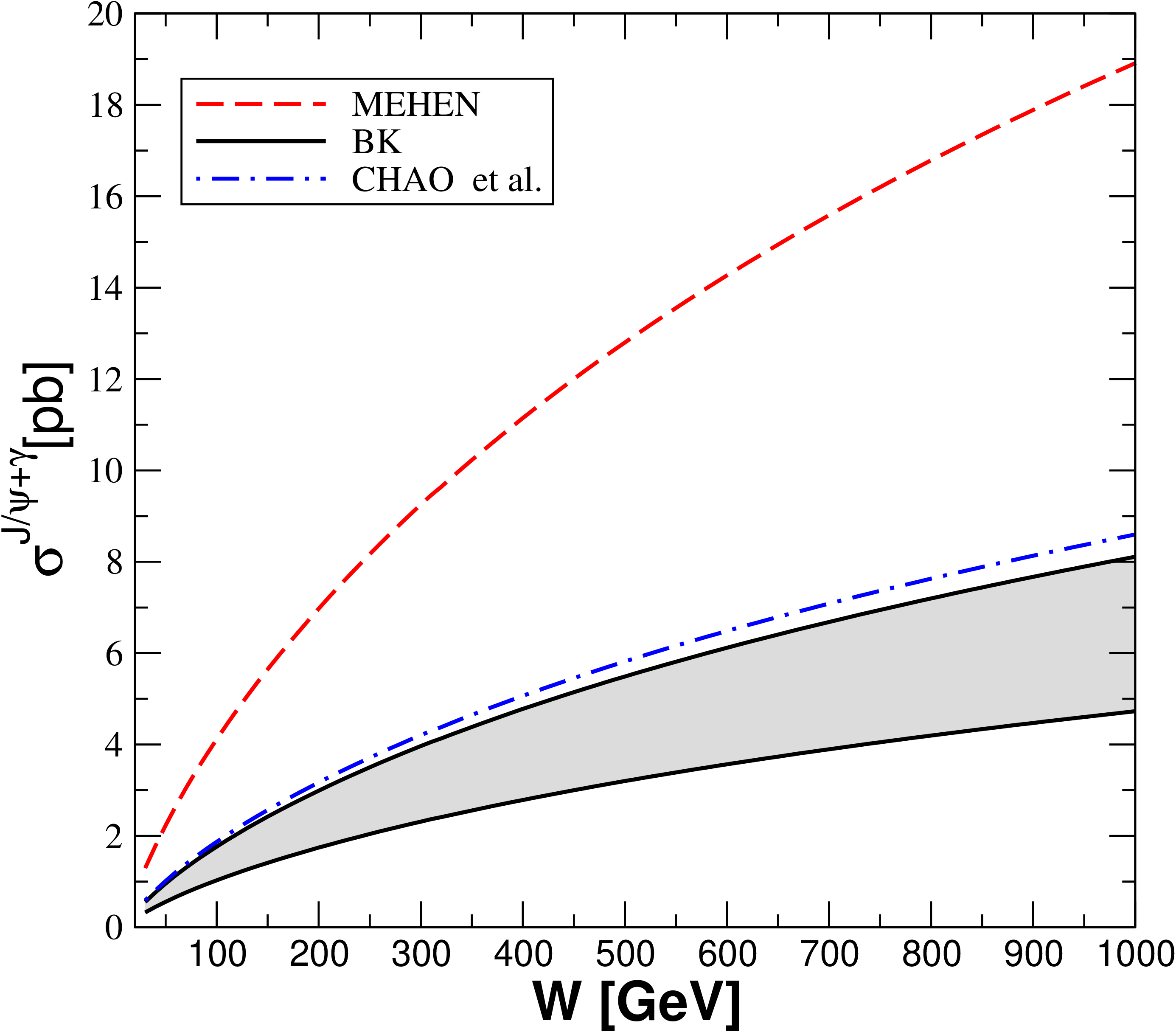}
\caption{Energy dependence of the total $J/\Psi + \gamma$ photoproduction cross section considering three different sets of matrix elements (see text).}
\label{fig1}
\end{figure}

\section{The Quarkonia+$\gamma$ Photoproduction }

  In the NRQCD formalism the cross section for the production of a heavy quarkonium state $H$ factorizes as  $\sigma (ab \rightarrow H+X)=\sum_n \sigma(ab \rightarrow Q\bar{Q}[n] + X) \langle {\cal{O}}^H[n]\rangle$, where the coefficients $\sigma(ab \rightarrow Q\bar{Q}[n] + X)$ are perturbatively calculated short distance cross sections for the production of the heavy quark pair $Q\bar{Q}$ in an intermediate Fock state $n$, which does not have to be color neutral.  The $\langle {\cal{O}}^H[n]\rangle$
are nonperturbative long distance matrix elements, which describe the transition of the intermediate $Q\bar{Q}$ in the physical state $H$ via soft gluon radiation. Currently, these elements have to be extracted in a global fit to quarkonium data as performed, for instance, in Ref. \cite{bk}. It is important to emphasize that the underlying mechanics governing heavy quarkonium production is still subject of intense debate (For a recent review see, e.g., Ref. \citep{review_nrqcd}).


In the specific case of the $H + \gamma$ photoproduction, the total cross section can be expressed as follows \cite{mehen}
\begin{eqnarray}
\sigma (\gamma + p \rightarrow H + \gamma + X) = \int dz dp_{\perp}^2 \frac{xg(x,Q^2)}{z(1-z)} \frac{d\sigma}{dt}(\gamma + g \rightarrow H + \gamma) \label{sigmagamp}
\end{eqnarray}
where $z \equiv (p_H.p)/(p_{\gamma}.p)$, with $p_H$, $p$ and $p_{\gamma}$ being the four momentum of the quarkonium, proton and photon, respectively. In the proton rest frame, $z$ can be interpreted as the fraction of the photon energy carried away by the quarkonium. Moreover, $p_{\perp}$ is the magnitude of the quarkonium three-momentum normal to the beam axis.
The partonic differential cross section $d\sigma/dt$ is given by \cite{mehen}
\begin{eqnarray}
\frac{d\sigma}{dt}(\gamma + g \rightarrow H + \gamma) = \frac{64 \pi^2}{3}\frac{e_Q^4 \alpha^2\alpha_s m_Q}{s^2}
\left(\frac{s^2s_1^2+t^2t_1^2+u^2u_1^2}{s_1^2t_1^2u_1^2}\right)\langle O_8^{V}(^3S_1)\rangle \label{dsdt}
\end{eqnarray}
where $e_Q$ and $m_Q$ are, respectively, the charge and mass of heavy quark constituent of the quarkonium. The Mandelstam variables can be expressed in terms of $z$ and $p_{\perp}$ as follows:

\begin{eqnarray}
s & = & \frac{p_{\perp}^2+(2m_Q)^2(1-z)}{z(1-z)}\,\,, \nonumber\\
t & = & - \frac{p_{\perp}^2+(2m_Q)^2(1-z)}{z}\,\,, \nonumber\\
u & = & - \frac{p_{\perp}^2}{1-z} \,\,.
\end{eqnarray}

Moreover, $s_1 = {s} - 4 m_Q^2$, $t_1 = {t} - 4 m_Q^2$,  $u_1 = u - 4 m_Q^2$ and the Bjorken variable $x$ can be expressed by
\begin{eqnarray}
x = \frac{p_{\perp}^2+(2m_Q)^2(1-z)}{W_{\gamma p}^2z(1-z)} \,\,,
\end{eqnarray}
	where $W_{\gamma p}$ is the photon - proton center-of-mass energy.
In our calculations we consider the gluon distribution as given by the CTEQ6 parametrization \cite{cteq}, $Q^2 = p_{\perp}^2+(2m_Q)^2$ and assume $m_c = 1.5$ GeV and $m_b = 4.5$ GeV.

\section{Results}
\label{results}

In what follows we present our results for the $H+\gamma$ photoproduction in coherent hadron - hadron collisions at LHC energies.
One of the main uncertainties in our predictions is associated to the NRQCD matrix elements. For the $J/\Psi$ case, we consider the more recent global fit performed in Ref. \cite{bk} (denoted BK hereafter), { and its uncertainties}, which obtain that $\langle O_8^{J/\psi}(^3S_1)\rangle=2.24 \pm 0.59 \times 10^{-3}$ GeV$^3$. For comparison, we also use the value previously considered in Ref. \cite{mehen} (denoted MEHEN hereafter):
$\langle O_8^{J/\psi}(^3S_1)\rangle=6.6 \times 10^{-3}$ GeV$^3$. Although this choice is outdated, { we keep it in order to estimate the dependence of the matrix elements in our results} for the $J/\Psi + \gamma$ production in coherent interactions. Moreover, we also consider the value recently obtained in Ref. \cite{Chao}: $\langle O_8^{J/\psi}(^3S_1)\rangle=3.0 \pm 1.2 \times 10^{-3}$ GeV$^3$ (denoted CHAO et al. hereafter). It is important to emphasize that the BK and CHAO et al. matrix elements are compatible within the errors. 
As show in Fig. \ref{fig1} the BK and MEHEN choices imply very distinct results for the $J/\Psi + \gamma$ photoproduction cross section, with the latter being an upper bound for the total cross section. As expected, our prediction using the central value for the CHAO et al. matrix element is very similar to the BK one. 
The cross section increases with the energy, which is directly associated to the $x$-behaviour of the gluon distribution. It is important to emphasize that while studies of photoproduction at HERA were limited to photon-proton center of mass energies of about 200 GeV, photon-hadron interactions at  LHC can reach one order of magnitude higher on energy. For instance,
if we consider
$pp$ collisions at LHC, the Lorentz factor  is
$\gamma_L = 7455$, giving the maximum c.m.s. $\gamma p$ energy
$W_{\gamma p} \approx 8390$ GeV. Consequently, the analysis can also be useful to constrain the proton gluon distribution.
(For related studies  see Ref. \cite{gluon}). We postpone a more detailed study of this topic for a forthcoming publication.

For the $\Upsilon$ case, we  use two different sets of values for the matrix elements. The first one taken from Ref. \cite{bk2} (denoted BSV hereafter), which obtain $\langle O_8^{\Upsilon}(^3S_1)\rangle=5.3\times 10^{-3}\pm 0.5$ GeV$^3$ from fits to CDF data for bottomonium production. The second one we taken from Ref. \cite{Braaten:2001} (denoted BFL hereafter) which obtain $\langle O_8^{\Upsilon}(^3S_1)\rangle=0.02$ GeV$^3$. Recently, the $\Upsilon$ prompt production at the Tevatron and LHC in NRQCD was studied in Ref. \cite{Chao2}. Unfornately, the fit of the Tevatron data performed in  \cite{Chao2} only determine linear combinations of the matrix elements, which implies that the exact value of  $\langle O_8^{\Upsilon}(^3S_1)\rangle$ is unknown \cite{Chao3}.

\begin{figure}
\begin{tabular}{cc}
\includegraphics[scale=0.15]{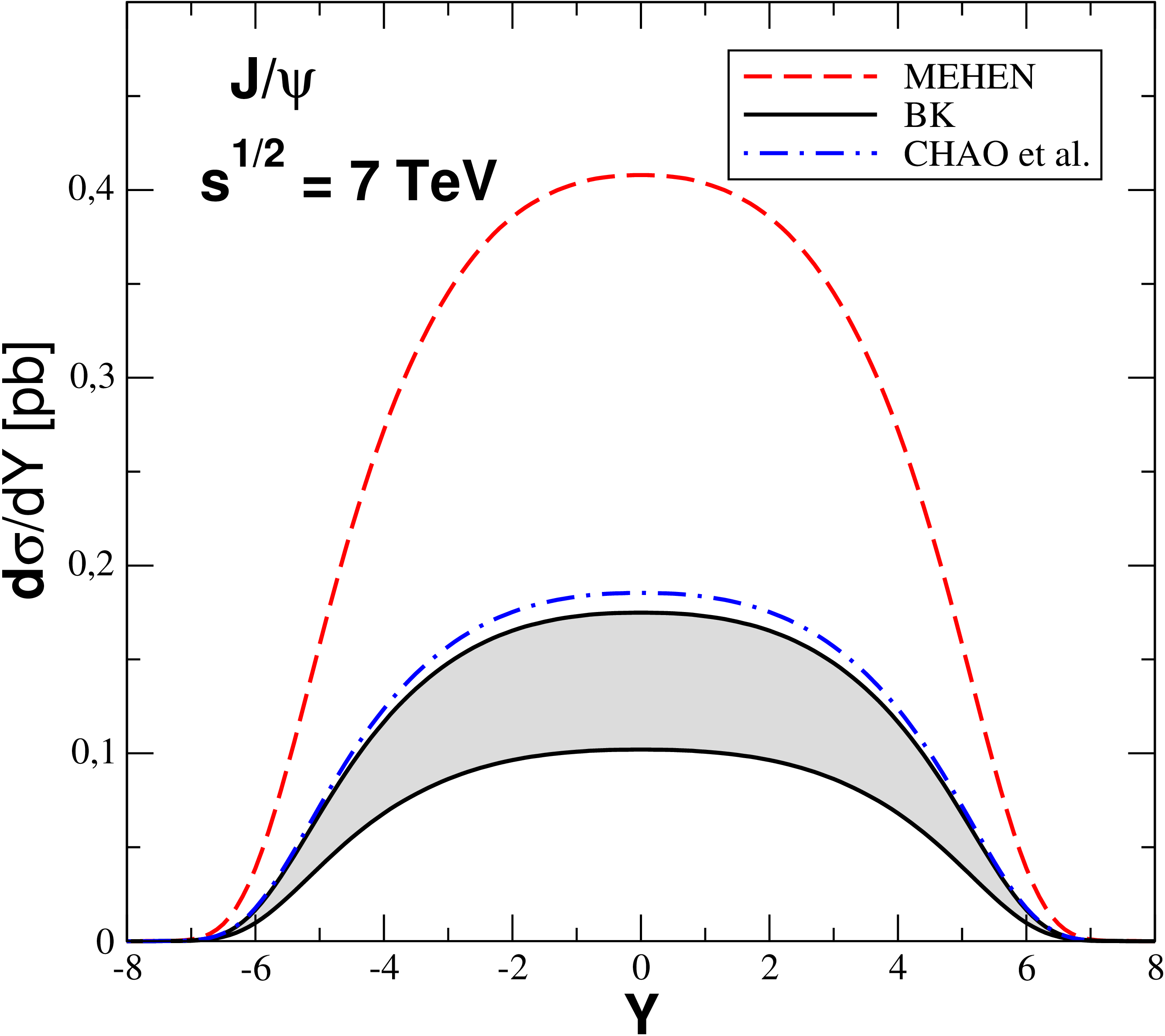} & \includegraphics[scale=0.15]{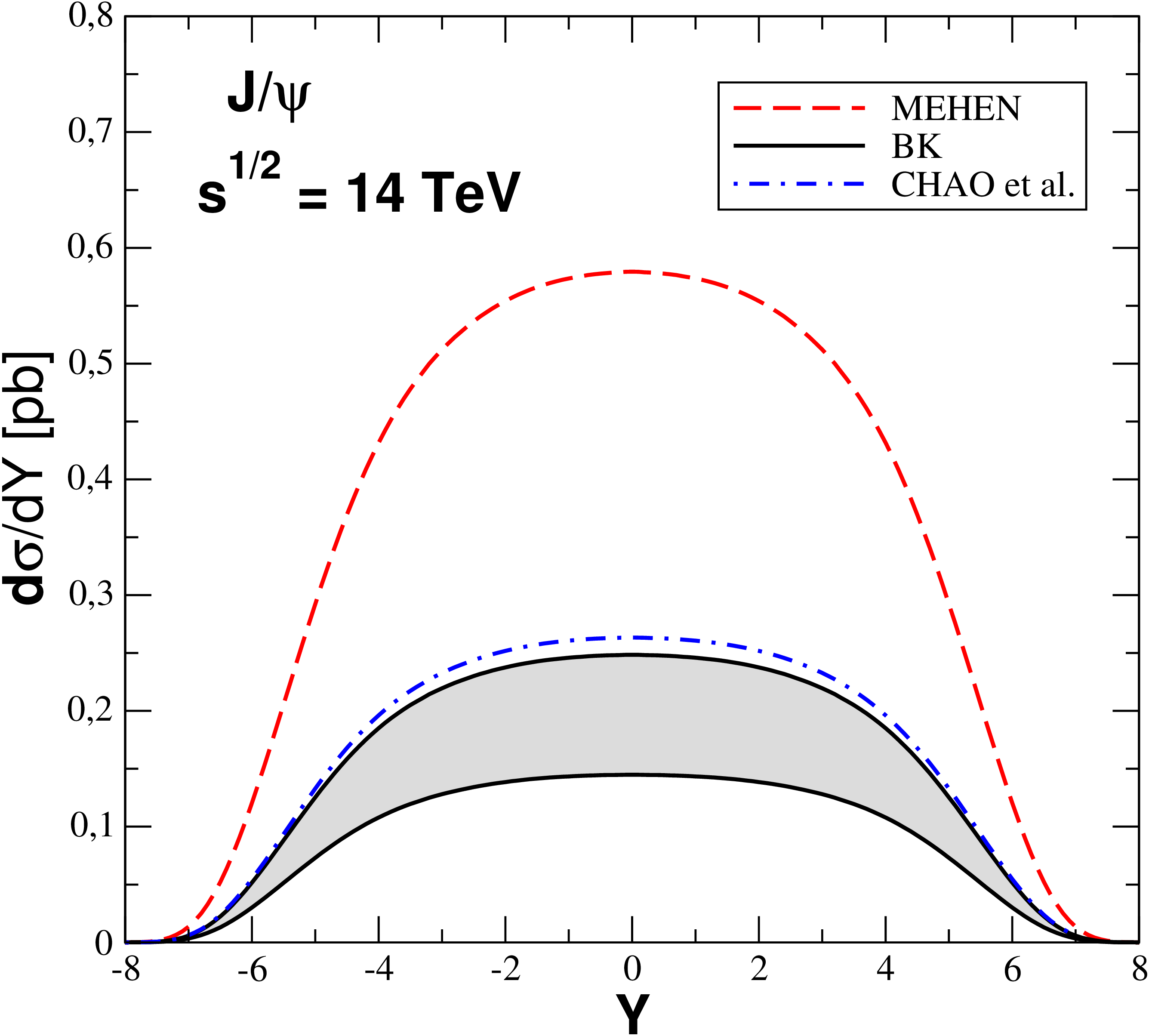}
\end{tabular}
\caption{Rapidity distribution for the $J/\Psi + \gamma$ photoproduction  in $pp$ collisions at $\sqrt{s} = 7$ TeV (left panel) and 14 TeV (right panel) considering three different sets of values for the matrix elements. }
\label{fig2}
\end{figure}

Lets now calculate the rapidity distribution and total cross sections for $H+ \gamma$  production in  coherent proton - proton  collisions at $\sqrt{s} = 7$ and 14 TeV.
The distribution on rapidity $Y$ of the produced final state can be directly computed from Eq. (\ref{sighh}), by using its  relation with the photon energy $\omega$, i.e. $Y\propto \ln \, (\omega/m_H)$.  Explicitly, the rapidity distribution is written down as,
\begin{eqnarray}
\frac{d\sigma \,\left[p+p \rightarrow p \otimes H + \gamma + X ) \right]}{dY} = \omega \, \frac{dN_{\gamma/h_1} (\omega )}{d\omega }\,\sigma_{\gamma h_2 \rightarrow H + \gamma + X}\,\left(\omega \right) + \omega \, \frac{dN_{\gamma/h_2} (\omega )}{d\omega }\,\sigma_{\gamma h_1 \rightarrow H + \gamma + X}\,\left(\omega \right)\,,
\label{dsdy}\end{eqnarray}
where we taken into account that the two protons can be the source of the photons ( $h_1 = h_2 = p$) and $X$ is a hadronic final state resulting of the  fragmentation of the proton target. The rapidity gap $\otimes$ is associated to the proton which emits the photon and remains intact.
In Fig. \ref{fig2} we present our results for the $J/\Psi + \gamma$ production considering the three different sets of values for the matrix elements and two different center-of-mass energies. As the photoproduction cross section is proportional to the cross section, our predictions for the rapidity distribution increases with the energy. As expected from Fig. \ref{fig1}, the predictions obtained using the  CHAO et al matrix element is similar to the BK one. Moreover, the predictions using the 
MEHEN matrix element  implies a value for rapidity distribution at $Y = 0$ which is a factor $\approx$ 3 larger than that using the BK one. This large  difference is also observed  in the predictions for the total cross section (See Table \ref{tab1}).
It demonstrates { that the} study of this observable can be useful to constrain the matrix element.
 Assuming the  design luminosity
${\cal L} = 10^7$ mb$^{-1}$s$^{-1}$ the corresponding event rates will be $\approx 1 \times 10^5$ ($2 \times 10^5$) events/years at $\sqrt{s} = 7 \, (14)$ TeV for the BK matrix elements.

\begin{figure}[t]
\includegraphics[scale=0.15]{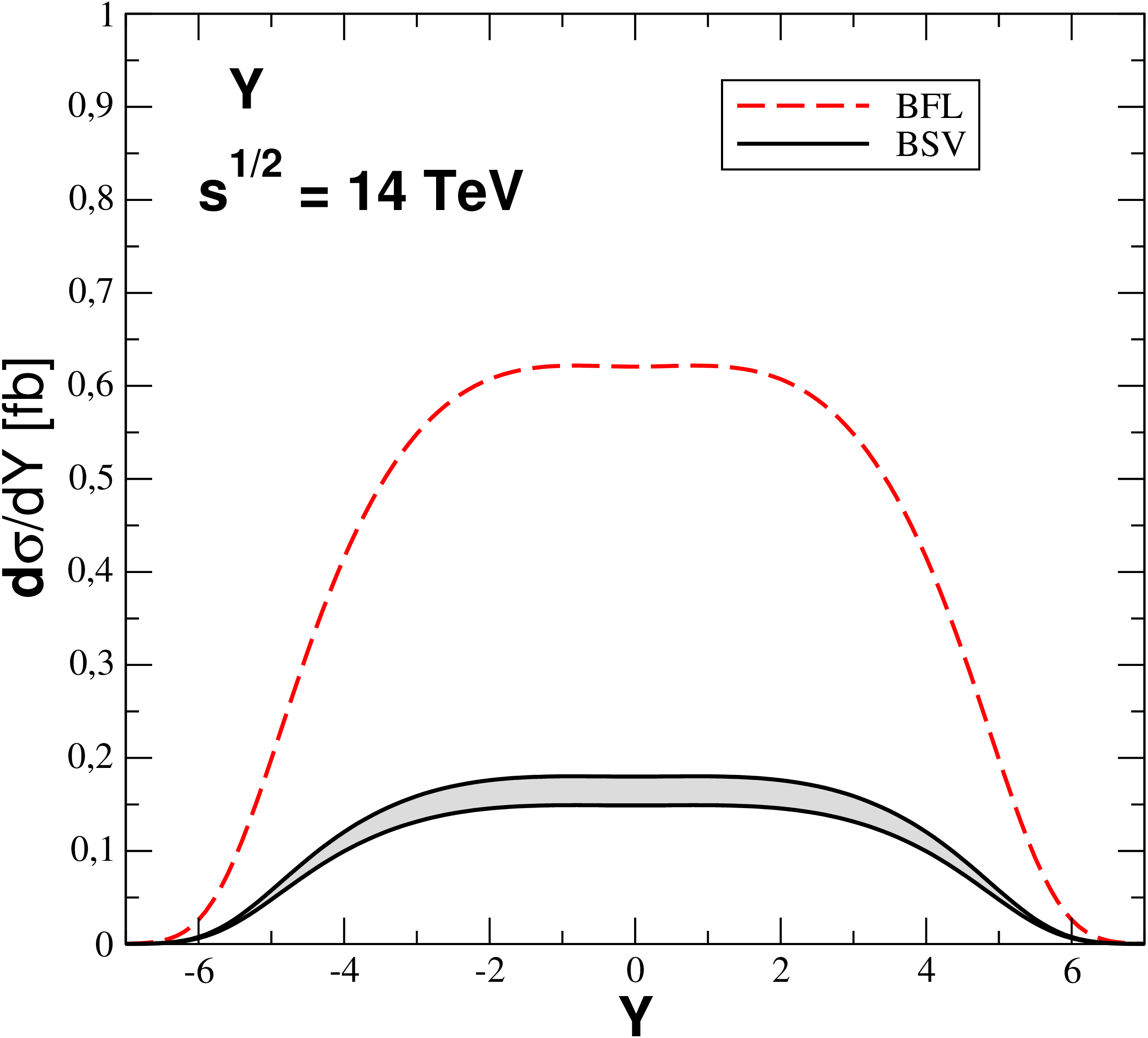}
\caption{Rapidity distribution for the $\Upsilon + \gamma$ photoproduction  in $pp$ collisions at $\sqrt{s} = 14$ TeV considering two different sets of values for the matrix elements. }
\label{fig3}
\end{figure}

In Fig. \ref{fig3} and Table \ref{tab1} we present our predictions for the $\Upsilon + \gamma$ production considering the two different sets of values for the matrix elements and $\sqrt{s} = 14$ TeV. As observed in the $J/\Psi$ case, the predictions are strongly dependent on the matrix elements used in the calculations. However, in the $\Upsilon$ case the cross sections are a factor $10^3$ smaller than those obtained for $J/\Psi + \gamma$ production, which implies that
the event rates will be $\approx 1 \times 10^2$  events/years, making the experimental analysis of this final state a hard task.

\begin{table}
\begin{center}
\begin{tabular} {||c|c|c||}
\hline
\hline
{\bf $J/\Psi + \gamma$} & {\bf MEHEN} & {\bf BK}   \\
\hline
\hline
LHC (7 TeV) & 3.62 pb   &  1.23 $\pm 0.50$ pb    \\
\hline
LHC (14 TeV) & 5.60 pb   &  1.90 $\pm 0.32$ pb    \\
\hline
\hline
{\bf $\Upsilon + \gamma$} & {\bf BFL}  & {\bf BSV} \\
\hline
\hline
LHC (14 TeV) & 5.46 fb   & 1.45 $\pm 0.13$ fb   \\
\hline
\hline
\end{tabular}
\end{center}
\caption{The total cross section  for the $H + \gamma$ photoproduction in coherent hadron - hadrons collisions at  LHC energies.}
\label{tab1}
\end{table}

\begin{figure}
\begin{tabular}{cc}
\includegraphics[scale=0.15]{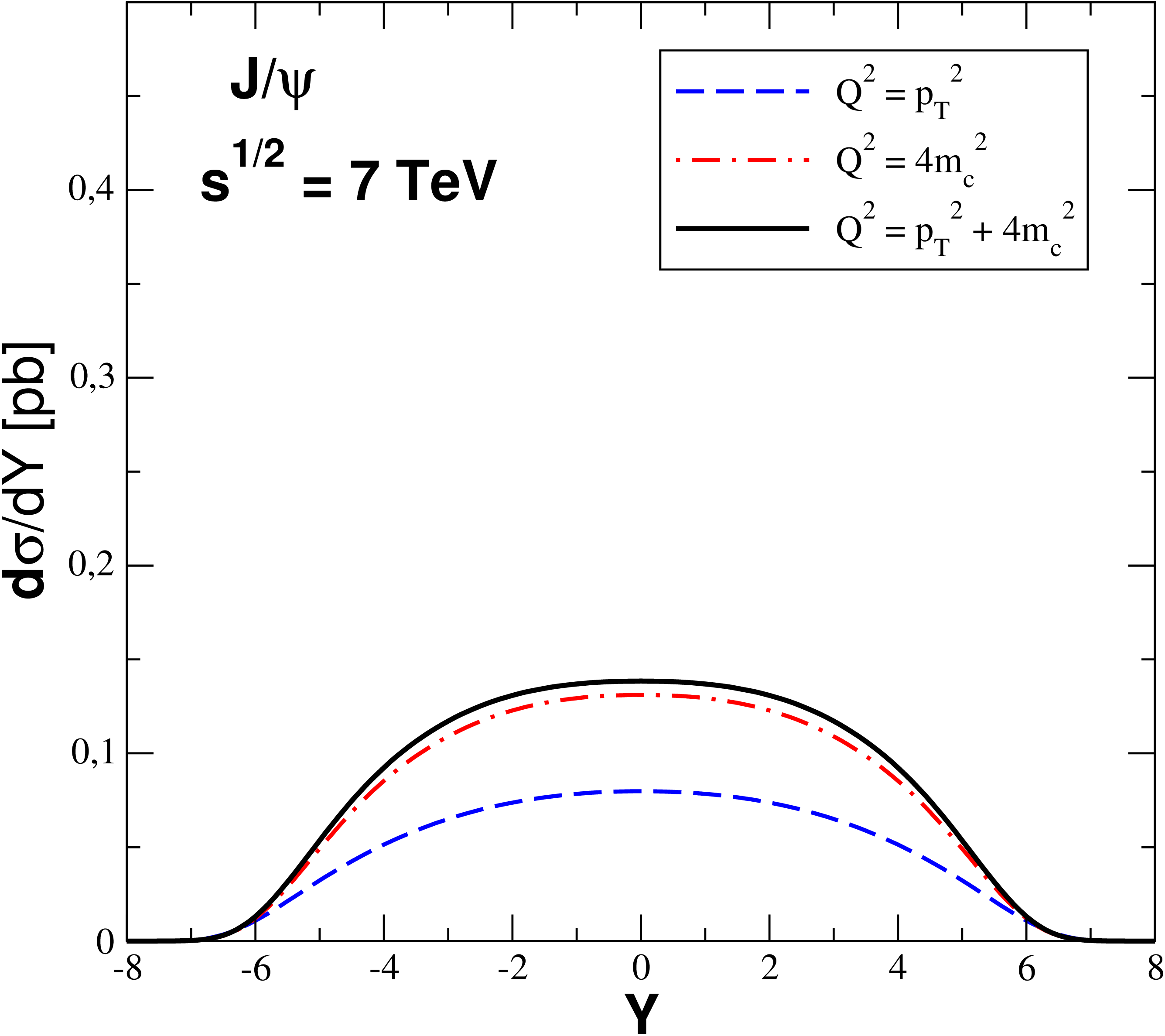} & \includegraphics[scale=0.15]{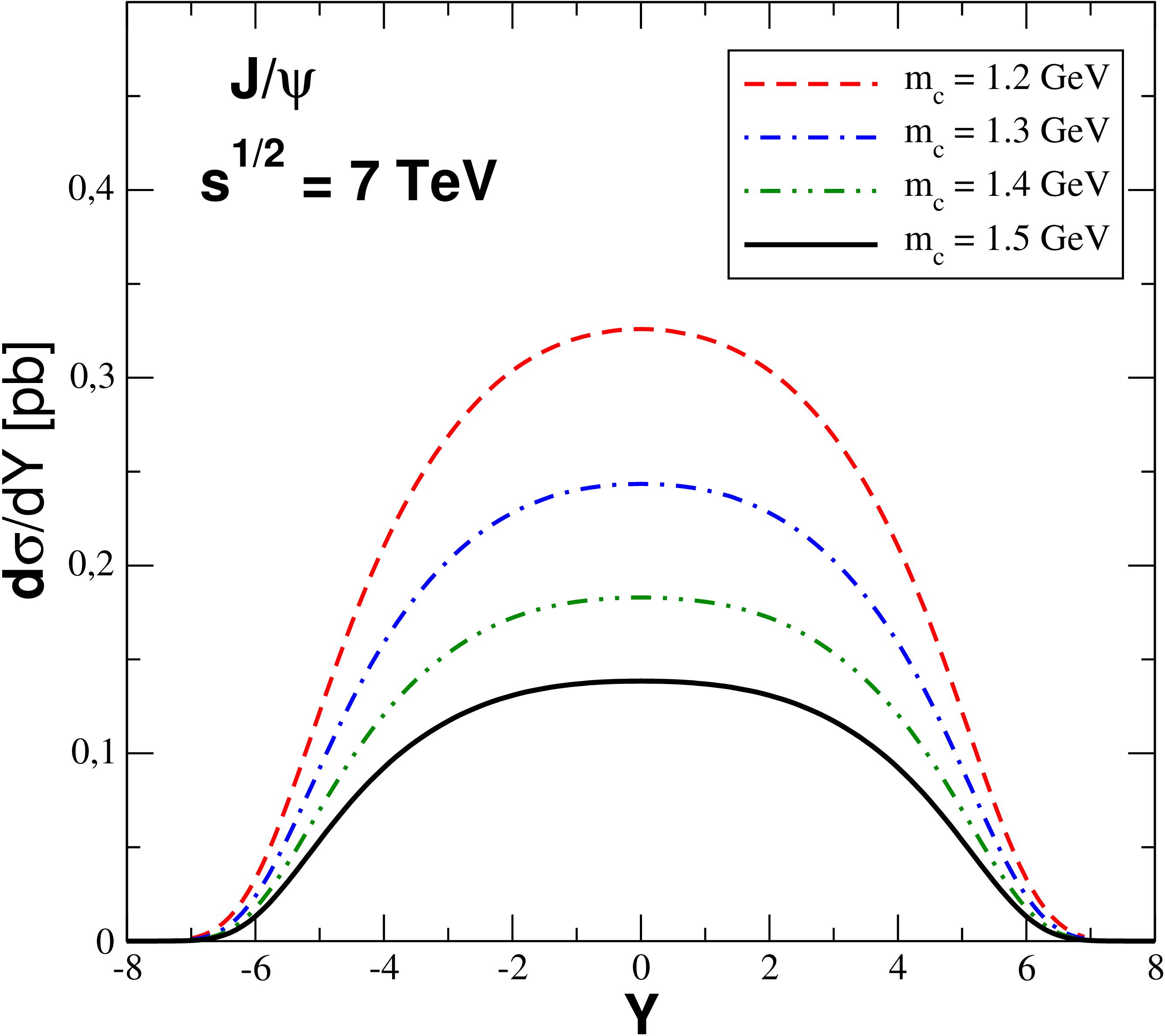}
\end{tabular}
\caption{Dependence on the  hard scale $Q^2$ (left panel) and charm mass (right panel)  of our predictions for the rapidity distribution for the $J/\Psi + \gamma$ photoproduction  in $pp$ collisions at $\sqrt{s} = 7$ TeV.}
\label{fig4}
\end{figure}

As demonstrated above, our predictions are strongly dependent on the matrix elements used in our calculations. Another uncertainty  comes from  on the choices for the hard scale $Q^2$ and the heavy quark mass. In the previous figures for the $J/\Psi + \gamma$ production we consider $Q^2 = p_{\perp}^2 + 4 m_c^2$ and $m_c = 1.5$ GeV. In Fig. \ref{fig4} we present our predictions considering other choices for $Q^2$ and $m_c$ and using the central value for the BK matrix element. In the left panel we assume $m_c = 1.5$ GeV and consider two other possibilities for the hard scale: $Q^2 = p_{\perp}^2$ and $Q^2 = 4m_c^2$. These new choices imply that the rapidity distribution at $Y = 0$ can be reduced by a factor two. In the   
right panel we assume $Q^2 = p_{\perp}^2 + 4 m_c^2$ and consider different values of the charm mass. Our predictions are strongly dependent on the value used in the calculation, with the rapidity distribution at $Y = 0$ increasing by a factor $\approx$ 3 if we consider $m_c = 1.2$ GeV. Similar dependences are verified in our predictions  for the $\Upsilon + \gamma$ production.

Finally, lets discuss the experimental separation of the  $H + \gamma$ photoproduction  and compare our predictions with those obtained   for the inclusive production or considering   diffractive (Pomeron) interactions. In comparison to the  inclusive hadroproduction (See e.g. \cite{Liwang}), which is characterized by the process $p + p \rightarrow X + H + \gamma + Y$, with both proton producing hadronic final states, the   photoproduction cross sections are a factor $\approx 10^3$ smaller. However, as in
 photoproduction we have a rapidity gap in the final state,  the separation of the signal from hadronic background would be relatively clear since the event multiplicity for photoproduction interactions is lower, which implies that it may be used as a separation factor between these processes.
Another mechanism which also is characterized by one rapidity gap in the final state, is the single diffractive $H + \gamma$ production. This process was analysed, e.g., in Refs. \cite{Xu,mmm},  considering that the Pomeron has a partonic structure. In comparison with those results, our predictions are a factor $\approx 8$ smaller.
However, it is expected that emerging hadrons from  Pomeron processes have a much larger transverse momentum than those resulting from  photoproduction processes. Consequently, in principle it is possible to introduce a selection criteria  to separate these two processes. Moreover, it is important to emphasize that Pomeron predictions are strongly dependent on the value used for the gap survival factor, while our results should not be modified by soft absorption corrections.


\section{Conclusions}
\label{conc}

In this paper we have computed for the first time the cross sections for photoproduction of quarkonium$+ \gamma$ in coherent $pp$ collisions at LHC energies using the NRQCD formalism and considering  different sets of values for the matrix elements.  Such  processes are interesting since the final state is characterized by a low multiplicity and one rapidity gap. Moreover, the   produced large $p_T$ quarkonia are relatively easy to detect through their leptonic decay modes and their transverse momenta are balanced by the associated high energy photon.
Our results demonstrate that the rapidity distributions and total cross sections are strongly dependent on the magnitude of the matrix elements. Moreover, we predict sizeable for the $J/\Psi + \gamma$ cross section, which makes the experimental analyses of this process feasible at LHC.

\section*{Acknowledgments}
This work was supported by CNPq, CAPES and FAPERGS, Brazil.

\end{document}